\documentclass[runningheads]{cl2emult}

\usepackage{makeidx}  
\usepackage{graphicx} 
\usepackage{multicol} 
\usepackage{cropmark} 
\usepackage{eso}      
\makeindex            

\newcommand{\euler}[1]{{\usefont{U}{eur}{m}{n}#1}}

\newcommand{\umu}{\mbox{\euler{\char22}}}

\usepackage{color}

\begin{document}
\title*{Near-Infrared Integral Field Spectroscopy of Damped Lyman-$\alpha$ Systems}
\toctitle{Near-Infrared Integral Field Spectroscopy of Damped Lyman-$\alpha$ Systems}
%
%
\titlerunning{Near-IR Integral Field Spectroscopy of DLAs}
%
\author{Andrew Bunker\inst{1}
\and Annette Ferguson\inst{1}
\and Rachel Johnson\inst{1}
\and Richard McMahon\inst{1}
\and Ian Parry\inst{1}
\and Max Pettini\inst{1}
\and Alfonso Arag\'{o}n-Salamanca\inst{2}
\and Rachel Somerville\inst{1}
}
\authorrunning{Andrew Bunker et al.}
%
%
\institute{Institute of Astronomy, Madingley Road, Cambridge CB3\,OHA, UK
\and School of Physics \& Astronomy, University of Nottingham, NG7\,2RD, UK
}

\maketitle              

\begin{abstract}
We assess the feasibility of detecting star formation in damped
Lyman-$\alpha$ systems (DLAs) at $z>1$ through near-infrared
spectroscopy using the forthcoming integral field units on 8\,m-class
telescopes. Although their relation to galaxies is not well established,
high-$z$ DLAs contain most of the neutral gas in the Universe, and this
reservoir is depleted with time -- presumably through star
formation. Line emission should be an indicator of star formation
activity, but searches based on Lyman-$\alpha$ are unreliable because of
the selective extinction of this resonant UV line. Using more robust
lines such as H$\alpha$ forces a move to the near-infrared at $z>1$. For
line emission searches, spectroscopy is more sensitive than imaging, but
previous long-slit spectroscopic searches have been hampered by the
likelihood that any star forming region in the DLA galaxy disk would
fall outside the narrow slit. The new integral field units such as
CIRPASS on Gemini will cover sufficient solid angles to intercept these,
even in the extreme case of large galactic disks at high redshift. On an
8\,m-class telescope, star formation rates of $<
1\,M_{\odot}\,{\rm yr}^{-1}$ will be reached at $z\approx
1.4$ with H$\alpha$ in the $H$-band. Such star formation rates are well
below $L^{*}$ for the high-$z$ Lyman-break population, and are
comparable locally to the luminous giant H{\scriptsize~II} complexes in
M\,101. It appears that instruments such as CIRPASS on Gemini will have
both the sensitivity and the survey area to measure star formation rates
in $z>1$ DLAs. These observations will probe the nature of damped
Lyman-$\alpha$ systems and address their relation to galaxies.
\index{abstract}
\end{abstract}

\section{Damped Systems and Galaxy Evolution}
Traditionally, flux-limited selection has been used to chart galaxy
evolution. The measurement of quasar absorption lines allows an
independent approach to studying the history of galaxies. The highest
hydrogen column density absorbers seen in the spectra of background QSOs
-- the damped Lyman-$\alpha$ systems (DLAs) -- contain most of the
neutral gas in the Universe at $z>1$ (ref.~1), although their exact
relation to galaxies is not well established.  The global history of
star formation can be inferred from the evolution in the co-moving
density of neutral gas (predominantly in DLAs) as it is consumed in star
formation (Fig.~1a \& refs.~2,3).  The average star formation rate in
each DLA depends then on their space density. One school of thought has
$z>2$ DLAs being thick gaseous disks, the progenitors of massive spirals
(e.g., ref.~4).  Alternatively, DLAs could be more numerous gas-rich
dwarfs, potentially sub-galactic building blocks (e.g., Fig.~1b and
refs.~5, 6\,\&\,7).

Hence, to determine the nature of the DLAs requires a measurement of
their star formation rate. This is most directly done through the
hydrogen recombination lines (e.g., ref.~8), which should have
luminosity proportional to the photo-ionizing flux produced by the most
massive and shortest lived OB stars (and therefore trace the
near-instantaneous star formation rate).  However, the largely
unsuccessful searches for Ly$\alpha$\,121.6\,nm emission do not provide
useful limits as this line is resonantly scattered and hence is
selectively suppressed by dust absorption, which is difficult or
impossible to quantify.  Moving to the Balmer lines
H$\alpha$\,656.3\,nm and H$\beta$\,486.1\,nm greatly reduces the
problem of dust extinction, enabling the true star formation rates to be
more accurately measured (e.g., ref.~9). For interesting redshifts where
large samples of DLAs exist ($z>1$), these lines are redshifted into the
near-infrared pass-bands.

\begin{figure}[h]
\centering
\resizebox{0.49\textwidth}{!}{\includegraphics{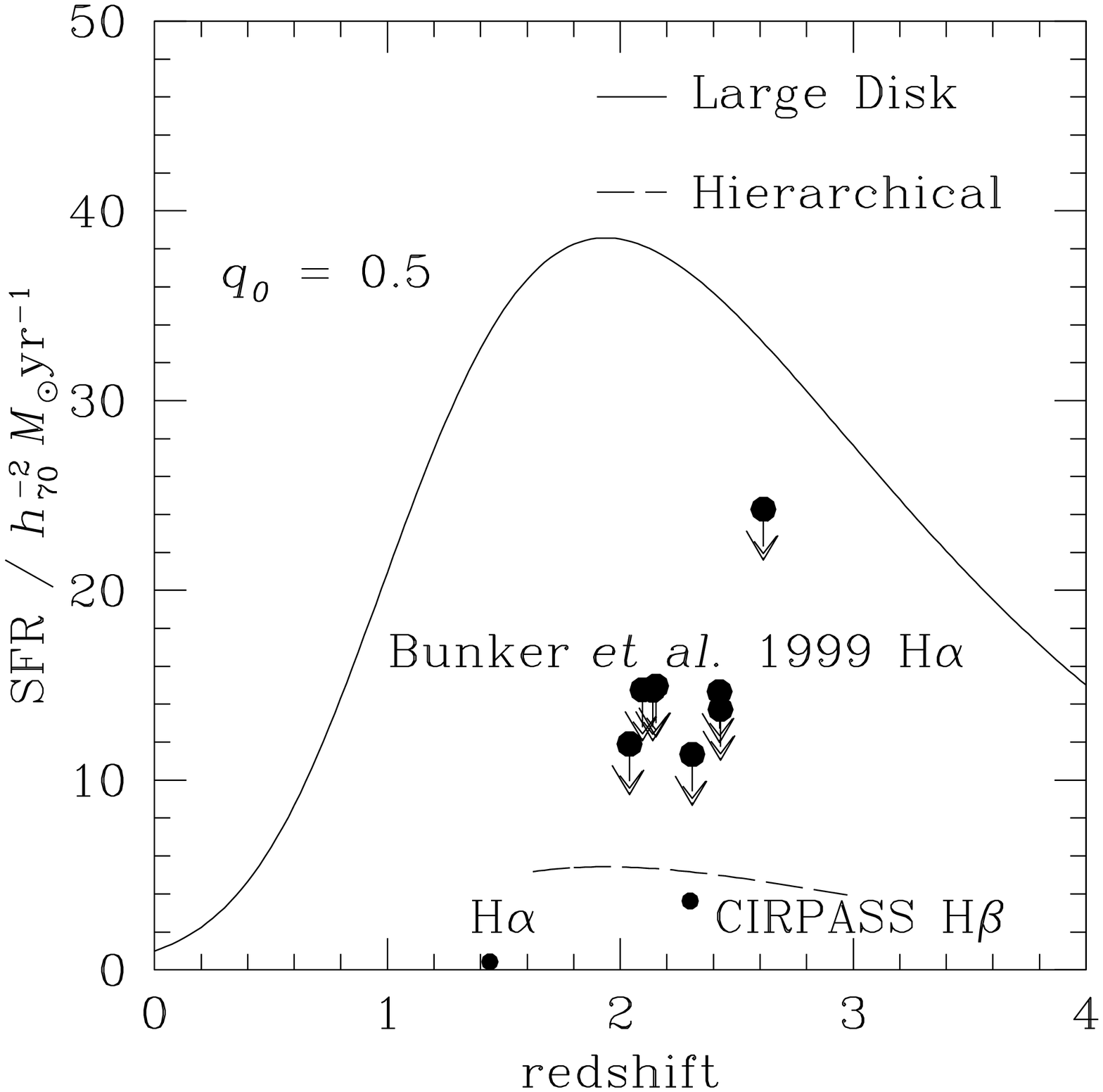}
}
\resizebox{0.49\textwidth}{!}{\includegraphics*[75,310][505,725]{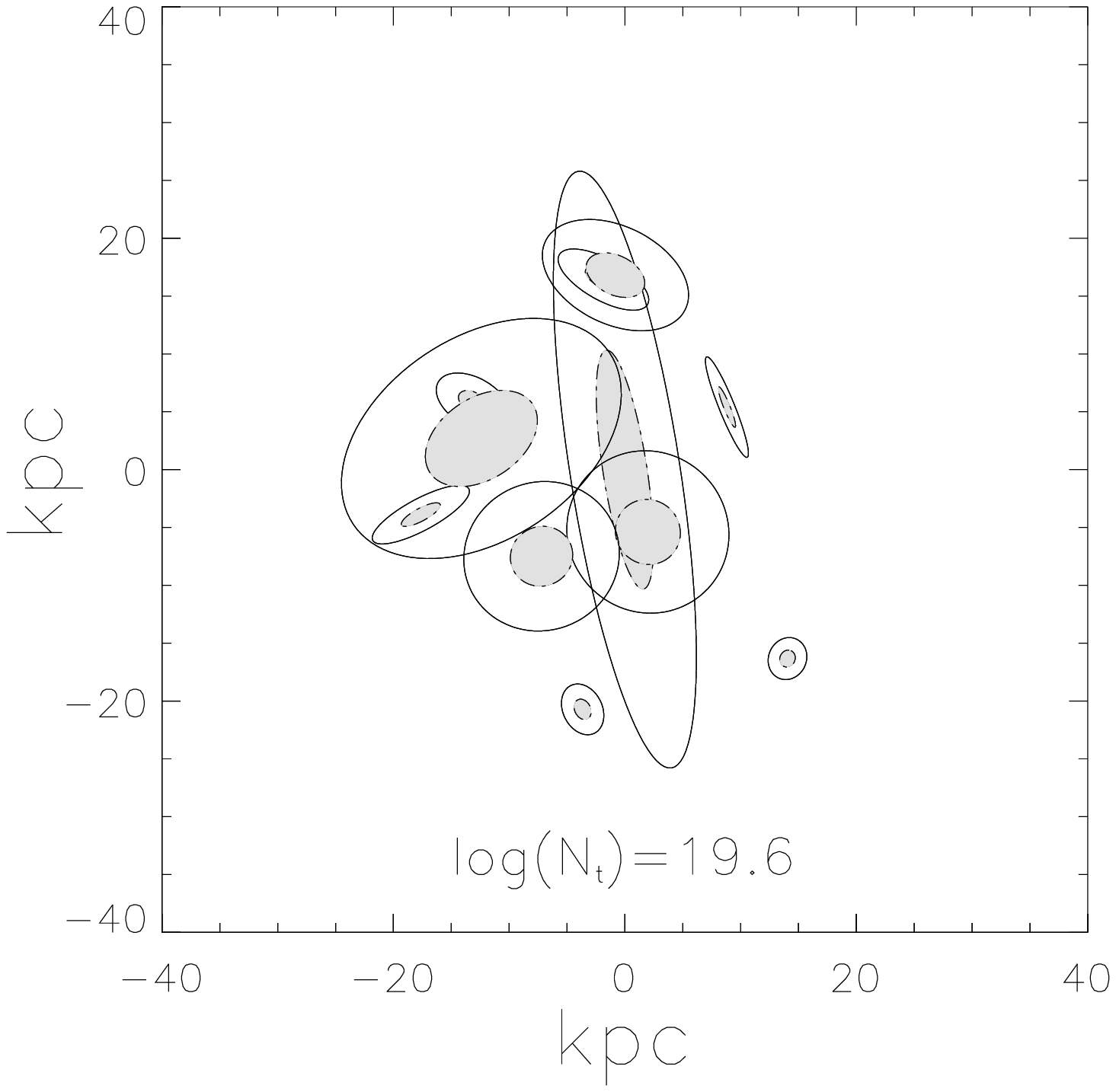}\hfil
\put(-140,100){\scalebox{0.4}{\includegraphics{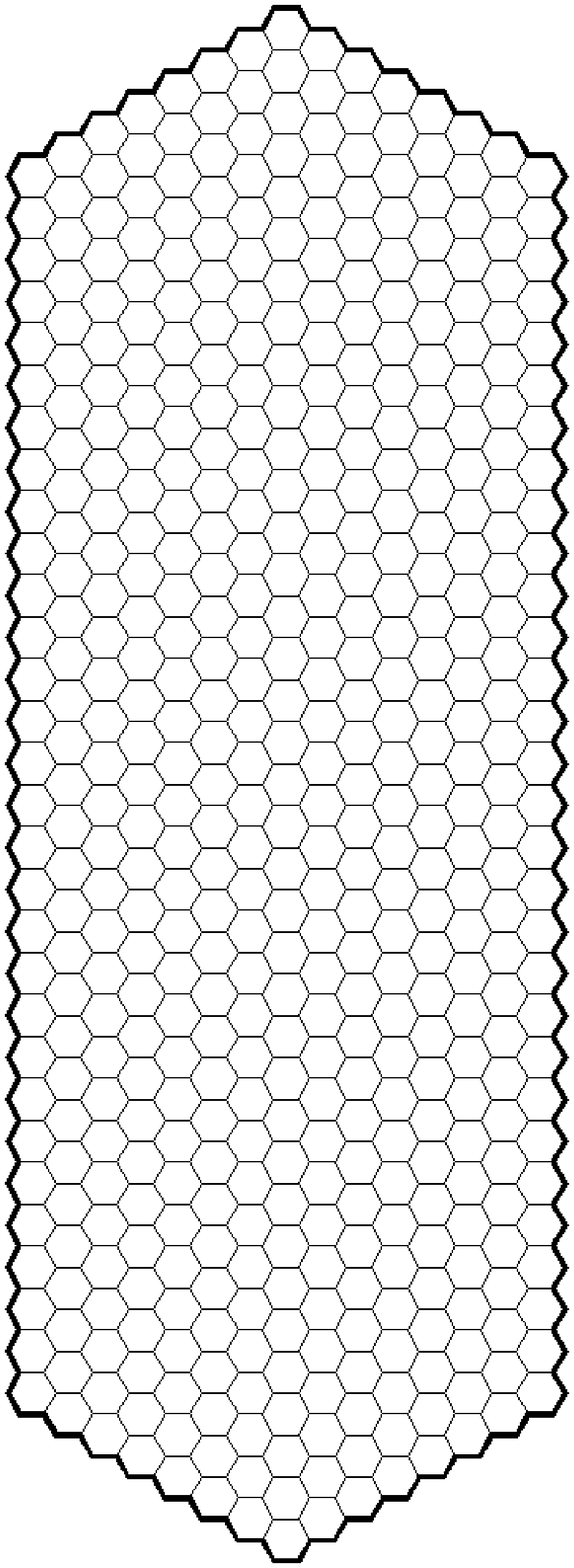}}}
}
\put(-300,143){\large\bf (a)}
\put(-130,142){\large\bf (b)}
\caption{{\bf (a)} The observational limits on the star formation rates
in high-reshifted damped Lyman-$\alpha$ systems, based on H$\alpha$
near-infrared long-slit spectroscopy with CGS\,4 on UKIRT (ref.~10). The
solid curve is the predicted average star formation rates for a
closed-box model (ref.~3) for the hypothesis where spirals today evolve
from high-redshift large disks.  The dashed curve is the
prediction if present day spirals are assembled from hierarchical
mergers of smaller, but more numerous, high-redshift DLAs (the
hierarchical hypothesis). The curves plotted take no account of the
possibility that the regions of star formation fall off the slit (which
may be $>90$\% of the cases using a spectroscopic slit).  Also shown are
the predicted sensitivities of the CIRPASS instrument on Gemini, which
will test {\em all} these models, unlike the existing observations. The
large area covered by the CIRPASS integral field unit means a much
greater likelihood of intercepting any star forming regions. \newline
{\bf (b)} A semi-analytic model of a halo with $V_{\rm
circ}=160$\,km\,s$^{-1}$ at $z=3$ (ref.~6).  The shaded region shows
the area of the disk with column density in excess of $2\times
10^{24}\,{\rm m}^{-2}$ (corresponding to DLAs). The overlaid CIRPASS
fibres (0.25\,arcsec across) shows that a single pointing of the
integral field unit centred on a DLA will intercept most of the high
column density gas -- the potential sites of star formation}
\label{fig:SFRs}
\end{figure}

\section{Exploring DLAs with Near-IR Spectroscopy}
Typically, we do not know where the star forming regions in the galaxy
associated with the DLA actually are -- all we know is that the
sight-line to the background QSO passes through a large column of
neutral hydrogen.  The difficulties of point spread function subtraction
of the QSO mean that broad-band imaging is unlikely to reveal the
foreground galaxy in emission at small impact parameters (separations
from the QSO sight line). However, spectroscopy is more effective for
line emission searches because it is more sensitive and any emission
line from the foreground DLA should be readily apparent against the
continuum of the background QSO.

\begin{figure}[h]
\centering
\resizebox{0.66\textwidth}{!}{\includegraphics{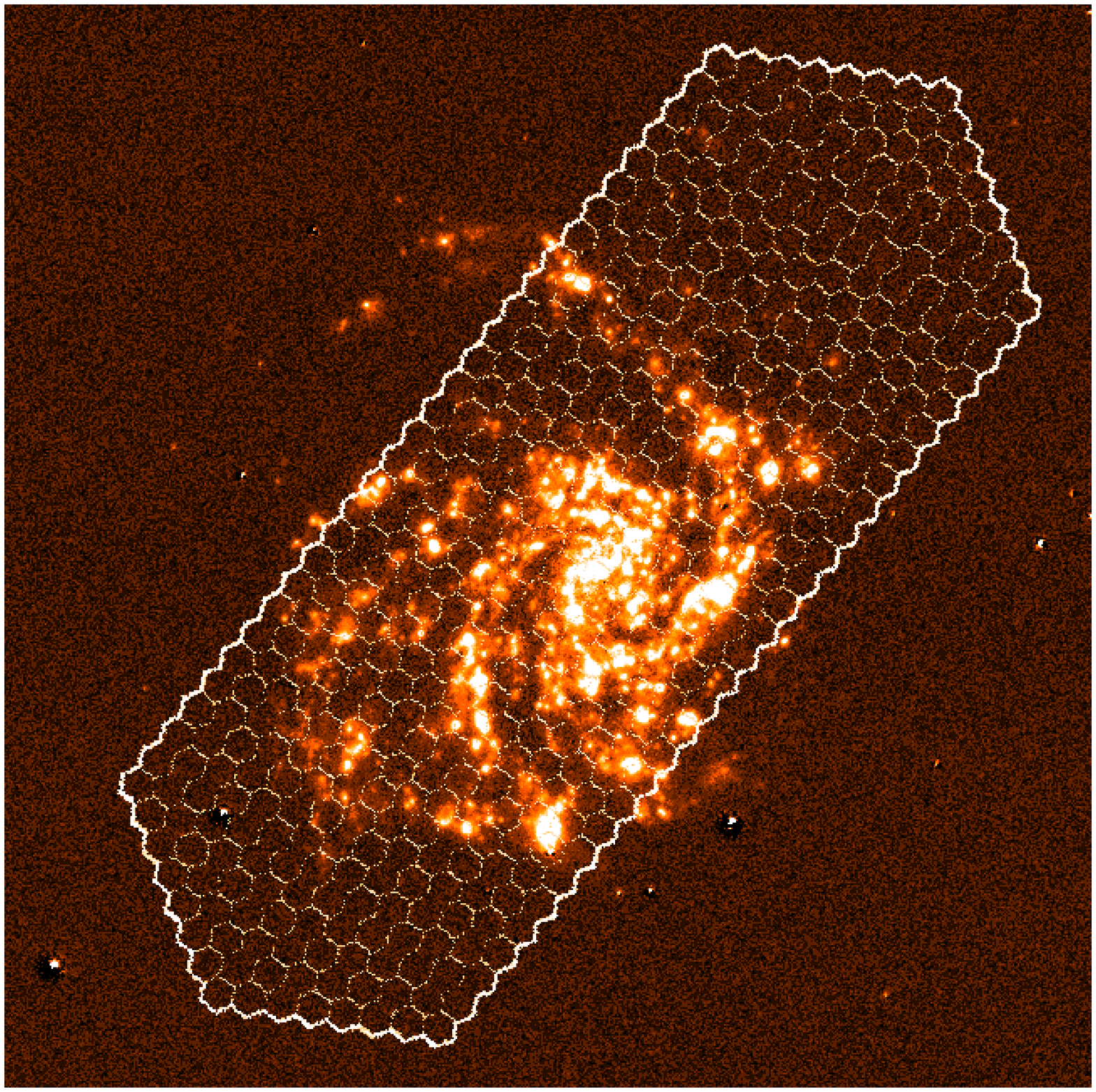}}
\resizebox{0.33\textwidth}{!}{\includegraphics{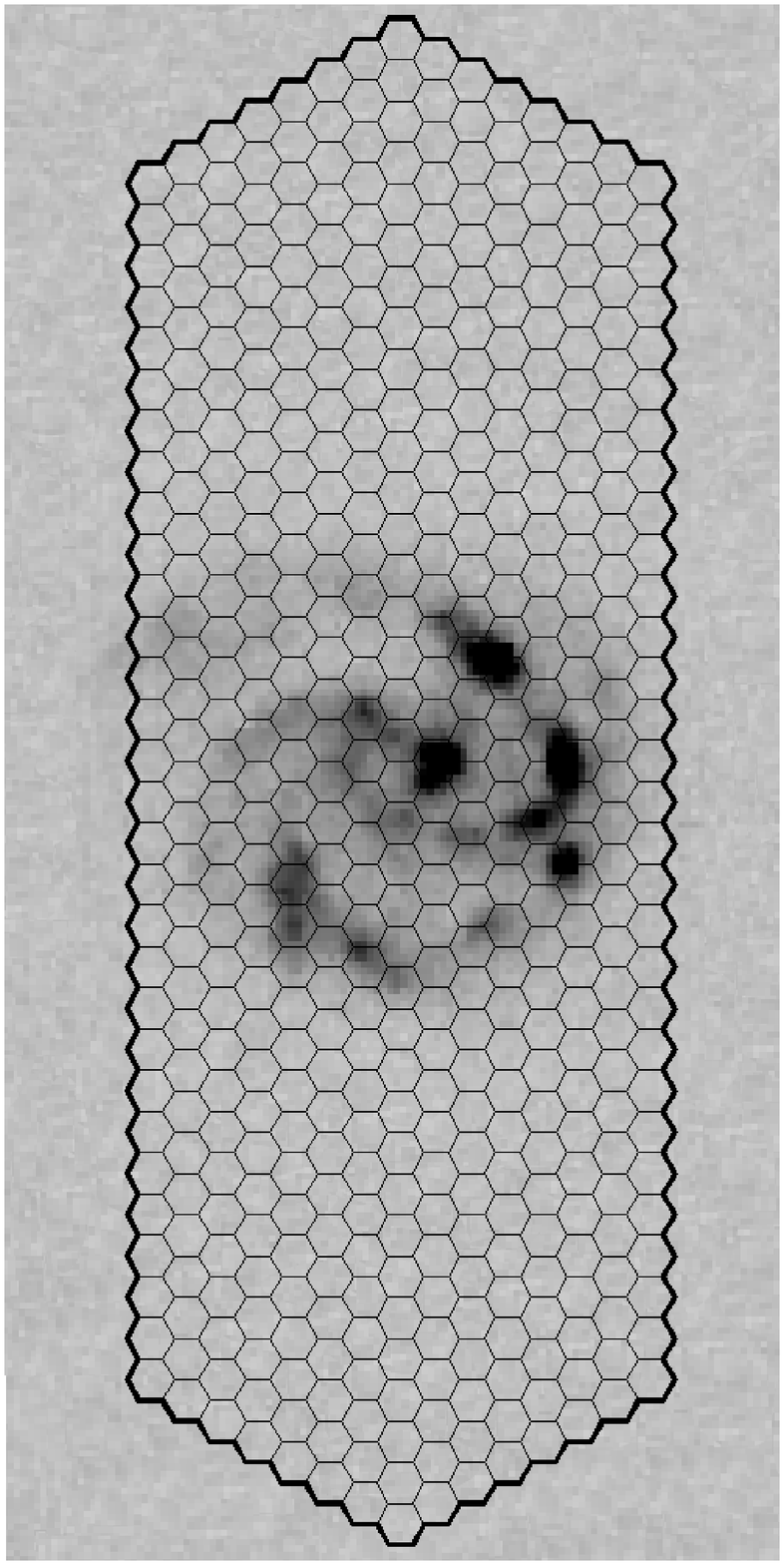}}
\put(-320,205){\textcolor{white}{\large\bf (a)}}
\put(-100,205){\large\bf (b)}
\caption{{\bf (a)} An H$\alpha$ image of the local spiral NGC\,4254 as
it would appear at $z=1.44$ with the CIRPASS integral field unit
overlaid (using 0.25\,arcsec diameter fibres). CIRPASS observations will
test whether $z>1$ DLAs really are large disk galaxies.
\newline
{\bf (b)} A spiral galaxy at $z\approx 1$ from the Hubble Deep Field
$B$-band.  The star-forming H{\scriptsize~II} regions are prominent in
the rest-frame UV. CIRPASS will accurately determine the true star
formation rates, since (1) the compact knots of star formation are
well-matched to the fibre size, reducing the sky background and
increasing the sensitivity; (2) the large area surveyed by the integral
field unit covers most of a spiral disk and (3) the H$\alpha$ line is a
much more robust measure of the star formation rate than the
dust-suppressed UV continuum and resonantly-scattered Lyman-$\alpha$}
\end{figure}

Recent near-infrared $K$-band long slit spectroscopy (ref.~10) with
CGS\,4 on the UKIRT 3.8\,m telescope failed to detect H$\alpha$ from
star formation in 8 DLAs at $z\approx 2.3$. However, the long-slit
approach may be ineffective, because of the high probability of any star
forming region in the DLA galaxy falling outside the narrow slit. If
DLAs really are $\sim L^{*}$ galaxies, then they could extend over tens
of kpc (several arcsec). Clearly, a larger solid angle needs to be
surveyed. This will be achieved with the new generation of integral
field units such as the Cambridge Infrared Panoramic Survey Spectrograph
(CIRPASS, ref.~11), which can cover $13''\times 4''$ with its 499 fibres.

Although the gaseous disks may be large, the star-forming
H{\scriptsize~II} regions are likely to be compact (as in local spirals
-- e.g., Fig.~2a). The small size of star forming regions is consistent
with observations of high-redshift galaxies in the rest-UV (Fig.~2b \&
ref.~12), and will be well-matched to the 0.25\,arcsec CIRPASS fibres
($\sim$\,1\,kpc across).  A 3-hour integration on one of the 8\,m Gemini
telescopes should achieve a 10$\sigma$ sensitivity between the sky lines
of $10^{-20}\,{\rm W\,m^{-2}}$ at $1.6\,\umu$m ($H$-band) for a compact
source, conservatively assuming that 70\% of the line emission from an
H{\scriptsize~II} region will fall within 4 fibres with seeing effects.
For a cosmology with ${\mathrm\Lambda}=0$, $H_{0}=70\,h_{70}\,{\rm
km\,s}^{-1}\,{\rm Mpc}^{-1}$ and $q_{0}=0.5(0.0)$, then the star
formation rates reached are $0.6\,(1.1)\,h_{70}^{-2}\,M_{\odot}\,{\rm
yr}^{-1}$ for H$\alpha$ at $z=1.4$ and
$5.1\,(14)\,h_{70}^{-2}\,M_{\odot}\,{\rm yr}^{-1}$ for H$\beta$ at
$z=2.3$ (see ref.~8).  Such star formation rates are well below $L^{*}$
for the high-$z$ Lyman-break population (e.g, refs.~9\,\&\,13) and are
comparable locally to the luminous giant H{\scriptsize~II} complexes in
M\,101.

It appears that instruments such as CIRPASS on Gemini will have both the
sensitivity and the survey area to measure star formation rates in $z>1$
DLAs. Coupled with studies of the evolution of neutral hydrogen (ref.~2)
and metal enrichment (ref.~14), these observations will probe the nature
of damped Lyman-$\alpha$ systems and address their relation to galaxies.

\clearpage
\addcontentsline{toc}{section}{Index}
\flushbottom
\printindex

\end{document}